\documentclass[11pt,a4paper]{article}
\usepackage{jheppub}

\usepackage{amssymb}
\usepackage{amsfonts}
\usepackage{amsbsy}
\usepackage[all]{xy}
\usepackage{amsmath}

\usepackage{upgreek}

\usepackage{amssymb,amscd}
\usepackage{mathrsfs}
\usepackage{amsmath,amsthm}

\usepackage{slashed}

\makeatletter
\DeclareFontFamily{OMX}{MnSymbolE}{}
\DeclareSymbolFont{MnLargeSymbols}{OMX}{MnSymbolE}{m}{n}
\SetSymbolFont{MnLargeSymbols}{bold}{OMX}{MnSymbolE}{b}{n}
\DeclareFontShape{OMX}{MnSymbolE}{m}{n}{
    <-6>  MnSymbolE5
   <6-7>  MnSymbolE6
   <7-8>  MnSymbolE7
   <8-9>  MnSymbolE8
   <9-10> MnSymbolE9
  <10-12> MnSymbolE10
  <12->   MnSymbolE12
}{}
\DeclareFontShape{OMX}{MnSymbolE}{b}{n}{
    <-6>  MnSymbolE-Bold5
   <6-7>  MnSymbolE-Bold6
   <7-8>  MnSymbolE-Bold7
   <8-9>  MnSymbolE-Bold8
   <9-10> MnSymbolE-Bold9
  <10-12> MnSymbolE-Bold10
  <12->   MnSymbolE-Bold12
}{}

\let\llangle\@undefined
\let\rrangle\@undefined
\DeclareMathDelimiter{\llangle}{\mathopen}%
                     {MnLargeSymbols}{'164}{MnLargeSymbols}{'164}
\DeclareMathDelimiter{\rrangle}{\mathclose}%
                     {MnLargeSymbols}{'171}{MnLargeSymbols}{'171}
\makeatother

\def\be{ \begin{equation} }
\def\ee{ \end{equation}}

\def\sp{\mathpzc{p}}
\def\sz{\mathpzc{z}}

\def\exp{{\rm exp}}

\def\Hom{{\rm Hom}}
\def\I{{\rm i}}

\renewcommand{\Im}{{\rm Im }}
\def\ker{{\rm ker}}

\def\half{\frac{1}{2}}
\def\p{\partial}

\def\one{{\hbox{ 1\kern-.8mm l}}}

\DeclareFontFamily{OT1}{pzc}{}
\DeclareFontShape{OT1}{pzc}{m}{it}{<-> s * [1.10] pzcmi7t}{}
\DeclareMathAlphabet{\mathpzc}{OT1}{pzc}{m}{it}

\def\CG {{\cal G}}
\def\CH {{\cal H}}

\def\CK {{\cal K}}
\def\CL {{\cal L}}
\def\CM {{\cal M}}

\def\CO {{\cal O}}

\def\CX {{\cal X}}
\def\CO {{\cal O}}

\def\CG {{\cal G}}
\def\CH {{\cal H}}

\def\CS {{\cal S}}

\def\CX {{\cal X}}
\def\CY {{\cal Y}}

\def\IC{\mathbb{C}}
\def\ID{\mathbb{D}}

\def\IR{{\mathbb{R}}}

\def\IZ{{\mathbb{Z}}}

\def\fg{\mathfrak{g}}

\def\fo{\mathfrak{o}}
\def\fp{\mathfrak{p}}

\def\fs{\mathfrak{s}}
\def\ft{\mathfrak{t}}

\def\fp{\mathfrak{p}}

\def\fs{\mathfrak{s}}
\def\ft{\mathfrak{t}}
\def\fu{\mathfrak{u}}

\def\rmG{{\mathrm{G}}}

\def\rmk#1{\bigskip\noindent{\bf Remarks} }

\newcommand\fro{{\overline{\underline{\Omega}}}}

\def\froM{ \overline{\underline{\CM}} }

\def\Dsl{\slashed{D}}

\title{$L^2$-Kernels Of Dirac-Type Operators On Monopole Moduli Spaces}

\author[a]{Gregory W.~Moore,}
\author[b]{Andrew B.~Royston,}
\author[c]{Dieter Van den Bleeken}

\affiliation[a]{NHETC and
Department of Physics and Astronomy, Rutgers University \\
126 Frelinghuysen Rd., Piscataway NJ 08855, USA}
\affiliation[b]{George P.\ \& Cynthia Woods Mitchell Institute for Fundamental Physics and Astronomy, \\
Texas A\&M University, College Station, TX 77843, USA}
\affiliation[c]{Physics Department, Bo\u{g}azi\c{c}i University\\
 34342 Bebek / Istanbul, TURKEY}

\emailAdd{gmoore@physics.rutgers.edu}
\emailAdd{aroyston@physics.tamu.edu}
\emailAdd{dieter.van@boun.edu.tr}

\abstract{We state some mathematical predictions concerning the kernels of
Dirac-type operators on moduli spaces of (singular) monopoles in $\IR^3$.
These predictions follow from the semiclassical interpretation of physical results
on spaces of (framed) BPS states in d=4, N=2 gauge theories.}

\begin{document}
\begin{flushright} MI-TH-1602 \end{flushright}
\maketitle

\section{Introduction And Conclusion}

  The purpose of this note is to
state some mathematical predictions for the differential
geometry of moduli spaces of monopoles that follow from
a semiclassical interpretation of some of the recent
results on the BPS spectrum of four-dimensional
gauge theories with N=2 supersymmetry.  These predictions concern the nature of
the $L^2$-kernel of a family of Dirac-type operators on moduli spaces
of both singular and non-singular monopoles. We state three conjectures
in Section \S \ref{sec:Predictions}.
The first is a generalization of the famous conjecture
of Ashoke Sen concerning the $L^2$-harmonic forms on monopole
moduli space \cite{SegalSelby,Sen:1994yi}. The second and third conjectures describe the
way in which the kernel is expected to jump as parameters in
the family are continuously varied. We illustrate the conjectures
with two examples in Section \ref{sec:Examples}. In this note we only consider ``pure''
vectormultiplet theories and ``pure'' 't Hooft line defects.
The generalizations to theories with matter and general
Wilson-'t Hooft line defects will appear elsewhere
\cite{Brennan}.

This note, aimed primarily at mathematicians,
 is a brief summary of the much more comprehensive paper
\cite{MRV3} where the reader can find extended material explaining
conventions, notation and background, as well as fuller explanations
of the reasoning leading to the statements we make here.
The papers \cite{Moore:2014jfa,Moore:2014gua,MRV3} also contain a more extensive list
of references.

\section{(Singular) Monopole Moduli Space}

Let $G$ be a compact simple Lie group. Consider Yang-Mills-Higgs
theory on $\IR^{1,3}$ with action proportional to
\be\label{eq:1}
  \int_{\IR^{1,3}}  (F,*F) + (DX,*DX).
\ee
Here $(\cdot, \cdot)$ is a Killing form, normalized so the long-root has
square-length $2$ and we have omitted the theta-term. (The latter is
fully taken into account in \cite{MRV3}.)  Magnetic monopoles are static
solutions of the Yang-Mills-Higgs equations defined by solutions of the Bogomolnyi equations
on $\IR^3$
\be\label{eq:2}
F = *_3 DX
\ee
satisfying the following boundary conditions: We choose a regular element
$X_\infty$ of the Lie algebra $\fg$ of $G$. This selects a Cartan subalgebra
$\ft$ as well as a system of simple roots $\alpha_I$, and coroots $H_I$,
where $I$ runs from one to the rank of the group. In a suitable gauge we
impose the asymptotic conditions as the radial distance $r$ from a choice of
spatial origin goes to infinity:
\be\label{eq:3}
\begin{split}
X  & = X_\infty - \frac{\gamma_{\mathrm{m}}}{2r} + \cdots \\
F  & = \half \gamma_{\mathrm{m}} \omega + \cdots \\
\end{split}
\ee
Here $\gamma_{\mathrm{m}}$ is  the magnetic charge and the gauge bundle, restricted to
spheres of constant radius $S^2_r$ for sufficiently large $r$, has a transition function on the
equator $e^{\I \phi} \rightarrow e^{\gamma_{\mathrm{m}} \phi}$,
so $\gamma_{\mathrm{m}} $ is in the coroot lattice
\footnote{We denote the coroot lattice by $\Lambda_{\mathrm{cr}}\subset \ft$. It is the integral dual
of the weight lattice $\Lambda_{\mathrm{wt}}\subset \ft^\vee$, and can be identified with
the elements $h\in \ft$ such that $\exp[2\pi h]=1$.  Similarly, the magnetic weight lattice
$\Lambda_{\mathrm{mw}} \subset \ft$ is the dual of  the root lattice of $\Lambda_{\mathrm{rt}} \subset \ft^\vee$.
In this paper we only consider adjoint-valued matter fields and hence we take the gauge group
to be the adjoint group. Thus the cocharacter lattice of $G$ is $\Lambda_{\mathrm{mw}}$ and the character
lattice of $G$ is $\Lambda_{\mathrm{rt}}$. Since we have chosen a Killing form on $\ft$, for any $h\in \ft$
we may define a dual vector $h^\vee\in \ft^\vee$ by $\langle h^\vee, h'\rangle := (h, h') $
for all $h'\in \ft$.  }
 $\Lambda_{\mathrm{cr}}$ of $\fg$, while
$\omega= \sin\theta d\theta d \phi$ is a volume form on $S^2_r$.
The moduli space of such solutions, identified by the group of gauge
transformations that approach $1$ as $r\to \infty$ is denoted  $\CM(\gamma_{\mathrm{m}}, X_\infty)$
and is
is a smooth hyperk\"ahler manifold \cite{Atiyah:1988jp}.
Expanding $\gamma_{\mathrm{m}} = \sum_I n_{\mathrm{m}}^I H_I$, with $n_{\mathrm{m}}^I\in \IZ$, the
moduli space is nonempty iff all $n_{\mathrm{m}}^I$ are nonnegative and at least one is positive \cite{Taubes:1981gw},
in which case the real dimension of the moduli space is $4 \sum_I n_{\mathrm{m}}^I$
\cite{Weinberg:1979ma,Weinberg:1979zt}.

There is a Lie algebra of Killing vector fields on $\CM(\gamma_{\mathrm{m}}, X_\infty)$ isomorphic to
$\IR^3 \oplus \fs\fo(3) \oplus \ft$, the summands corresponding to symmetries of translation, rotation about
the origin, and global gauge transformations, respectively. We define the
Killing vector associated with an element $h\in \ft$ as follows. In a gauge $A_0=0$
let $\hat A = A_i dx^i + X dx^4$ be a four-dimensional gauge field so that solutions of  \eqref{eq:2}
are equivalent to self-dual gauge fields $\hat A$ that are translation invariant in $x^4$.
Given such a solution solve
$\hat D^2 \epsilon =0$ for a smooth map $\epsilon: \IR^4 \to \fg$ that is translation
invariant in $x^4$ and satisfies $\epsilon \to h$ for $r\to \infty$. Then the
Killing vector $\rmG(h)$ has directional derivative at $\hat A$ given by
$\frac{d}{ds} \hat A = - \hat D \epsilon$. The map
$h \to \rmG(h)$ is a Lie algebra homomorphism from $\ft$ to the Lie algebra of triholomorphic
Killing vectors of the universal cover $\widetilde{\CM}(\gamma_{\mathrm{m}}, X_\infty)$  of $\CM(\gamma_{\mathrm{m}}, X_\infty)$. In this
paper we make the simplifying
assumption that $n^I_{\mathrm{m}} > 0$ for all $I$ so that the action is effective. (The more general
situation is fully treated in \cite{MRV3}.)  The universal cover is metrically a product
\be\label{eq:4}
\widetilde{\CM}(\gamma_{\mathrm{m}}, X_\infty) = \IR^4 \times \CM_0(\gamma_{\mathrm{m}}, X_\infty)
\ee
where the $\IR^4$ factor has the standard Euclidean metric
and is generated by the translation Killing vectors along with $\rmG(X_\infty)$.
The space $\CM_0(\gamma_{\mathrm{m}}, X_\infty)$ is the
``strongly centered''  moduli space in the terminology of \cite{HitchinMantonMurray}.
The moduli space is a quotient $\CM(\gamma_{\mathrm{m}}, X_\infty) =\widetilde{\CM}(\gamma_{\mathrm{m}}, X_\infty)/\ID$
where the group of Deck transformations $\ID \cong \IZ$, and is generated by an isometry $\phi$. An important point
for us is that the group of hyperholomorphic isometries acting on the universal cover and defined by
$\ID_{\rm g} := \{ \exp[2\pi \rmG(h)] \vert  h \in \Lambda_{\mathrm{mw}}\}$,
is in general a proper subgroup of $\ID$. Through any point $m_0\in \CM(\gamma_{\mathrm{m}}, X_\infty)$
there is an orbit of the maximal torus defining a homomorphism $ \pi_1(T,1) \to \pi_1(\CM,m_0)$.
We may view this as a homomorphism $\mu: \Lambda_{\mathrm{mw}} \to \IZ$. Based on the rational map formulation of monopole
moduli spaces \cite{Atiyah:1988jp,Donaldson,Hurtubise,Jarvis} we prove  that
$\mu(\lambda) = (\gamma_{\mathrm{m}}, \lambda)$  and   for any $\lambda\in \Lambda_{\mathrm{mw}}$
\be\label{eq:EqualIsom}
\exp[ 2\pi \rmG(\lambda)]  = \phi^{\mu(\lambda) }.
\ee
The image of $\mu$  is a subgroup $\ell \IZ \subset \IZ$
where $\ell$ is the positive integer such that
$\ell^{-1}\gamma_{\mathrm{m}}^\vee\in \Lambda_{\mathrm{rt}}$ is primitive.
(Equivalently, it is the gcd of the components of $\gamma_{\mathrm{m}}^\vee$
along a basis of simple roots.)

We now allow singularities in the monopole field corresponding, physically, to the introduction
of an 't Hooft line defect at the origin. The data of the singularity is given by
(the Weyl orbit of) an element $P$ of the cocharacter lattice $\Lambda_G = \Hom(U(1),T)$
where $T$ is the Cartan torus of $G$ with Lie algebra $\ft$. Singular monopoles satisfy equation \eqref{eq:2},
together with boundary conditions \eqref{eq:3} for $r\to \infty$ along with the boundary
conditions that read, in a suitable gauge,
\be\label{eq:5}
\begin{split}
X & = - \frac{P}{2r} + \CO(r^{-1/2}) \\
F & = \half P \omega + \CO(r^{-3/2} ) \\
\end{split}
\ee
for $r\to 0$. The space of solutions to \eqref{eq:2},\eqref{eq:3},and \eqref{eq:5},
quotiented by gauge transformations approaching $1$ at $r\to \infty$ and the normalizer
of $P$ for $r\to 0$, is the moduli space $\froM(P,\gamma_{\mathrm{m}}, X_\infty)$. It is a hyperk\"ahler
manifold with singularities on codimension four loci. The singularities are due to the phenomenon
of monopole bubbling \cite{Kapustin:2006pk,Moore:2014gua} and in some cases are loci of orbifold singularities.  Let
\be\label{eq:6}
\tilde \gamma_{\mathrm{m}} := \gamma_{\mathrm{m}} - P^- = \sum_I \tilde n^I_{\mathrm{m}} H_I
\ee
where $P^-$ is the Weyl image of $P$ in the chamber with $\langle \alpha_I, P \rangle \leq 0$ for all $I$.
It is conjectured in \cite{Moore:2014jfa} (with some supporting evidence given in \cite{Moore:2014gua})
that the moduli space is nonempty iff $\tilde n^I_{\mathrm{m}} \geq 0$. The presence of the singularity
at the spatial origin reduces the Lie algebra of Killing vectors to $ \fs\fo(3) \oplus \ft$.
The space $\froM(P,\gamma_{\mathrm{m}}, X_\infty)$ only depends on the Weyl group orbit $[P]$ of $P$,
up to hyperk\"ahler isometry.

\section{Consequences of $N=2$ Supersymmetry}

\subsection{General Remarks}

We now consider the above Yang-Mills-Higgs theory in the context of N=2 supersymmetric
gauge theory. The requirement of N=2 supersymmetry brings in a number of new features.
(See \cite{Weinberg:2006rq} for a comprehensive review.)

The first new feature is the presence of a second adjoint Higgs field $Y$.
 The Hamiltonian contains a potential energy term proportional to
$\int_{\IR^3} ([X,Y], * [X,Y])$ and this term forces finite energy field configurations to
obey  $Y \to Y_\infty$ for $r\to \infty$ with $Y_\infty \in \ft$.

A second new feature is the
existence in the quantum theory of distinguished quantum mechanical states known as
``BPS states'' \cite{Witten:1978mh,Seiberg:1994rs}. These are quantum mechanical analogs of the
classical dyon field configurations satisfying the Bogomolnyi bound, and indeed the latter
configurations are semiclassical approximations to BPS states at weak coupling. On the
Coulomb branch of vacua of the N=2 quantum field theory the gauge group is broken down
to the maximal torus $T$ and the BPS states carry electromagnetic charges with respect
to this unbroken gauge symmetry. In the semiclassical limit there is a distinguished
duality frame and the electromagnetic charge $\gamma$ is valued in the lattice
\footnote{Actually, the semiclassical regime is not simply connected, and
$\Gamma$ is really a local system. This well-known complication is thoroughly
discussed in \cite{MRV3}.}
\be\label{eq:7}
\Gamma := \Lambda_{\mathrm{mw}} \oplus \Lambda_{\mathrm{wt}} \subset \ft \oplus \ft^\vee.
\ee
The lattice $\Gamma$ has a natural symplectic form defined by
\be\label{eq:8}
\llangle \gamma_1, \gamma_2 \rrangle =
\llangle \gamma_{1,\mathrm{m}} \oplus \gamma^{\mathrm{e}}_1 , \gamma_{2,\mathrm{m}} \oplus \gamma^{e}_2 \rrangle:=
\langle \gamma^{\mathrm{e}}_1 , \gamma_{2,\mathrm{m}} \rangle - \langle \gamma^{e}_2 , \gamma_{1,\mathrm{m}} \rangle.
\ee
It is valued in $\frac{1}{\sz} \IZ$, where $\sz$ is the order of the center of the universal cover $\widetilde{G}$.

A third new feature is that the 't Hooft line defects can be generalized to line defects preserving four out of the
eight supersymmetries.  In the physical theory the line defect depends on a choice of a further parameter,
namely a phase $\zeta$ that determines which half-dimensional subspace of supersymmetries is preserved by the
line defect. Since this parameter will not be visible in our mathematical predictions we will simply denote
the line defect by $L([P])$.    The insertion of a line
defect changes the Hilbert space, but, when singularities at the origin are properly taken into account one can still
define a Bogomolnyi bound. The so-called \emph{framed BPS states} are the BPS states in the presence of a supersymmetric
line defect \cite{Gaiotto:2010be}. They have electromagnetic charge $\gamma$ in the $\Gamma$-torsor
$\left( P + \Lambda_{\mathrm{cr}}\right) \oplus \Lambda_{\mathrm{rt}}$.

A fourth new feature arising from N=2 supersymmetry is that low energy excitations of (singular) monopoles
(the dynamics of ``collective coordinates'') are described
by  supersymmetric quantum mechanics on moduli space  
\cite{Manton:1981mp,Gauntlett:1993sh,Gauntlett:1995fu,Gauntlett:1999vc,Weinberg:2006rq}.
The paper \cite{MRV3} has an in-depth review of these results, including generalizations accounting for the effects of
a theta angle, the presence of singular monopoles, and the inclusion of one-loop effects.

\subsection{Geometrical Definitions Of BPS States}

The supersymmetric dynamics of collective coordinates leads to
geometrical definitions of the spaces of BPS states.
For brevity we henceforth denote $\CX:= X_\infty$ and $\CY:=Y_\infty$.

We begin with the framed BPS states.
Let $\Dsl^\CY$ be the Dirac-type operator on the bundle of $L^2$-normalizable Dirac spinors
on $\froM(P,\gamma_{\mathrm{m}}; \CX)$ defined by $\Dsl^\CY = \Dsl + \slashed{\rmG}(\CY)$, where $\Dsl$ is the
standard Dirac operator and $\slashed{\rmG}(\CY)$ is Clifford multiplication by the one-form
derived from the triholomorphic Killing vector $\rmG(\CY)$.
For brevity, in what follows we will simply refer to $\Dsl^\CY$ as a
``Dirac operator.''

\bigskip
\noindent
\textbf{Definition}:   \emph{The   space of semiclassical framed BPS states, in the
vacuum $(\CX,\CY)$,  in the presence of the supersymmetric
defect $L([P])$, and which have   magnetic charge $\gamma_{\mathrm{m}} \in P + \Lambda_{\mathrm{cr}}$,  is: }
\be\label{eq:9}
 \ker_{L^2}(\Dsl^\CY ).
\ee

The reason for the phrase ``in the vacuum $(\CX,\CY)$'' is explained
in Remark 2 below.
In the semiclassical description the magnetic charge of the BPS states is encoded in the
choice of moduli space $\froM(P,\gamma_{\mathrm{m}}; \CX)$ whereas the electric charge is the
character of the   triholomorphic action of $\ft$ on $\ker_{L^2}(\Dsl^\CY )$.
Indeed, we can define the spaces of framed BPS states of fixed electromagnetic
charge $\gamma = \gamma_{\mathrm{m}} \oplus \gamma^{\mathrm{e}}$ from the
isotypical decomposition with respect to the action of $\ft$:
\be\label{eq:10}
 \ker_{L^2}(\Dsl^\CY ) = \oplus_{\gamma^{\mathrm{e}} \in \Lambda_{\mathrm{wt}}}
\overline{\underline{\CH}}([P]; \gamma;\CX,\CY).
\ee

A parallel definition can be given for the ordinary (``unframed'' or ``vanilla'') BPS states,
but there are a two extra complications since one must  ``factor out the center of mass degrees of freedom''
and then impose the proper equivariance condition with respect to the Deck group $\ID$.

As preparation introduce the rank one projection operator $\textbf{P}: \ft \to \ft$
onto the line spanned by $\CX$ and defined by
$\textbf{P}(h):=  \frac{(h,\gamma_{\mathrm{m}})}{(\CX,\gamma_{\mathrm{m}})}  \CX$.
Since $\textbf{P}$ is a projection operator
we have a direct sum decomposition:
\be\label{eq:DrctSum}
\ft \cong {\rm Im}(\textbf{P}) \oplus {\rm Im}(1-\textbf{P}) :=  \ft_{\mathrm{com}} \oplus \ft_0
\ee
of vector spaces.  The subspace $\ft_0$ is the subspace orthogonal to $\gamma_{\mathrm{m}}$ in the
Killing metric (but $\textbf{P}$ is \underline{not} the orthogonal projection in the Killing
metric). In \cite{MRV3} we prove that for all $h\in \ft$,
\be
(\rmG(h), \rmG(\CX)) = (h, \gamma_{\mathrm{m}}) .
\ee
On the left hand side of this equation we use the hyperk\"ahler metric on moduli space,
and on the right hand side we use the Killing metric on $\ft$.
Therefore, the orthogonal decomposition, using the  hyperk\"ahler metric,  along $T\IR^4 \oplus T\CM_0$
for the Killing vectors  $\rmG(h)$ takes the form
\be
\rmG(h) = \rmG(h_{\mathrm{com}}) + \rmG(h_0)
\ee
where, for any $h\in \ft$ we define $h_{\mathrm{com}}:= \textbf{P}(h)$
and $h_0 := (1-\textbf{P})(h)$.
Now we can choose orthogonal coordinates $(x^i, x^4)$ for the
$\IR^4$ factor so that the three translation Killing vectors
give translations in $x^i$ (holding all else fixed) and
$\frac{\p}{\p x^4}$ is a vector field parallel to $\rmG(\CX)$,
normalized so that
\be\label{eq:x4-norm}
(\frac{\p}{\p x^4}, \rmG(\CX))  := 1
\ee
It follows from \eqref{eq:EqualIsom} that $\phi^*(x^4) = x^4 + 2\pi$.

We now split the Dirac operator $\Dsl^\CY$ on
$\widetilde{\CM}(\gamma_{\mathrm{m}},X_\infty)$  using the splitting
of the tangent bundle following from \eqref{eq:4}:
\be\label{eq:SplitDirac}
\Dsl^{\CY} = \Dsl^{\CY}_{\rm com} + \Dsl^{\CY}_0.
\ee
Explicitly, in terms of the coordinates on $\IR^4$ defined above we have
\be\label{eq:Dcm}
\Dsl_{\mathrm{com}}^{\CY} = \Gamma^i \frac{\p}{\p x^i} + \Gamma^4 \left( \frac{\p}{\p x^4} - \I \frac{(\CY,\gamma_{\mathrm{m}})}{(\CX,\gamma_{\mathrm{m}}) } \right)
\ee
where $\Gamma^i,\Gamma^4$ are $4\times 4$ gamma matrices.
 We now define subspaces of sections of spinor bundles:
\be\label{eq:CKcm}
\CK_{\mathrm{com}}:= \ker \Dsl_{\mathrm{com}}^{{\rm Y}' } \subset \Gamma[\CS\to \IR^4]
\ee
\be\label{eq:CK0}
\CK_{0}:= \ker_{L^2} \Dsl_0^{{\rm Y}}  \subset \Gamma_{L^2}[\CS\to \CM_0]
\ee
with ${\rm Y}' \in \ft_{\mathrm{com}}$ and ${\rm Y}\in \ft_0$.
By pulling back from the two factors and multiplying we get a subspace of sections of
the spinor bundle on  $\widetilde{\CM}$:
\be\label{eq:KerSp}
\CK:= \pi_{\mathrm{com}}^*(\CK_{\mathrm{com}}) \otimes \pi_0^*(\CK_0)
\ee
and $\CK$ is annihilated by $\Dsl^{\CY}$ for $\CY = {\rm Y}' + {\rm Y}$.
There are no $L^2$-harmonic spinors on $\IR^4$ for $\Dsl^{\CY}_{\rm com}$.
This is physically reasonable since the monopoles are particles and definite
momentum eigenstates are plane waves.  On the other hand, we do not
wish to have such continuous degrees of freedom ``internally''
so it is important to impose the $L^2$ condition in the definition
of $\CK_0$.

Now, $\CK$ is a representation of $\ft$ and  we can decompose it into eigenspaces
of charge $\gamma^{\mathrm{e}} \in \ft^\vee$. Let $\CG(h)$ denote the
lift of the diffeomorphism $\exp[2\pi \rmG(h)]$ to the spinor bundle. (It exists.)
By definition, a ``spinor of electric charge $\gamma^{\mathrm{e}}\in\ft^\vee$''
is one such that for all  $h\in \ft$:
\be\label{eq:ElecCharge-Def}
\CG(h)^* \Psi = \exp[ -2\pi \I \langle \gamma^{\mathrm{e}}, h \rangle] \Psi.
\ee

The general element $\Psi_{\mathrm{com}}\in \CK_{\mathrm{com}}$  is of the form
$\Psi_{\mathrm{com}} = e^{-\I q x^4} s$, where  $s\in \IC^4$ is a constant Dirac spinor and
$q  =  -(\CY,\gamma_{\mathrm{m}})/(\CX,\gamma_{\mathrm{m}}) $.
On the other hand, the $\ft$-action on $\CK$ factors through a product of a $\ft_{\mathrm{com}}$ action on
$\CK_{\mathrm{com}}$ and a $\ft_0$-action on $\CK_0$. It follows that we must also have
\be\label{eq:Defq}
q  = \frac{\langle \gamma^{\mathrm{e}}, \CX \rangle}{( \gamma_{\mathrm{m}},\CX )}
\ee
and hence
\be\label{eq:Compat}
(\CY, \gamma_{\mathrm{m}}) + \langle \gamma^{\mathrm{e}}, \CX \rangle = 0
\ee
is a necessary condition for the existence of a spinor in $\CK$ with the data
$(\CX,\CY,\gamma_{\mathrm{m}},\gamma^{\mathrm{e}})$.

A spinor $\Psi\in \CK$ will only descend to $\CM(\gamma_{\mathrm{m}},\CX)$ if $\gamma^{\mathrm{e}}$ is properly quantized,
that is, if $\gamma^{\mathrm{e}} \in \Lambda_{\mathrm{mw}}^\vee = \Lambda_{\mathrm{rt}}$. However, because of
equation \eqref{eq:EqualIsom}  we must impose a further invariance condition
under a cyclic group $\IZ/\ell\IZ$. Thus the physically relevant kernel of the Dirac
operator is
\be\label{eq:DefVanillaBPS}
\left( \oplus_{\gamma^{\mathrm{e}} \in \Lambda_{\mathrm{rt}}}  \CK^{\gamma^{\mathrm{e}}}\right)^{\IZ/\ell\IZ}
\ee
and since $\phi$ commutes with the $\ft$ action we may impose the cyclic group invariance
on each isotypical summand. We define \eqref{eq:DefVanillaBPS} to be
\emph{ the space of semiclassical BPS states in the
vacuum $(\CX,\CY)$, with magnetic charge $\gamma_{\mathrm{m}}$.}
Since $\CK_{\mathrm{com}}\cong \IC^4$ we may factor it out of
 \eqref{eq:DefVanillaBPS} and define both the space of BPS states and the ``internal BPS states'' by
\footnote{When $\fs\fo(3)_{\rm rot} \oplus \fs\fu(2)_R$ are properly
defined $\CK_{\mathrm{com}}$ is the   $(\half;0)\oplus (0;\half)$ representation
and is known in physics as the ``half-hypermultiplet.''   }
\be\label{eq:BPSdef}
(\CK^{\gamma^{\mathrm{e}}})^{\IZ/\ell\IZ} := \CH(\gamma;\CX,\CY) := \CK_{\mathrm{com}} \otimes \CH_0(\gamma; \CX, \CY).
\ee

We conclude this section with three remarks

\begin{enumerate}

\item The  Killing vector fields arising from rotational symmetry around
the origin are not triholomorphic. Nevertheless, they preserve the $L^2$-kernel
of   $\Dsl^{\CY}$ and $\Dsl^{\CY}_0$ on $\overline{\underline{\CM}}$ and $\CM_0$
respectively  and hence the spaces
$\overline{\underline{\CH}}([P]; \gamma;\CX,\CY)$
and $\CH_0(\gamma; \CX, \CY)$ are representations of $\fs\fo(3)$.
Moreover, the commutant  $\fs\fu(2)$, in $\fs\fo(4N)$, of the $\fs\fp(N)$ holonomy
(where we denote the real
dimension of the moduli space $\overline{\underline{\CM}}$ or $\CM_0$ by $4N$) has a lift to the spinor bundle
and again acts on the kernel of $\Dsl^{\CY}$. There is therefore an action of
$\fs\fo(3)_{\rm rot} \oplus \fs\fu(2)_{R}$ on the BPS spaces. The subscripts indicate
that the symmetry arises from rotational symmetry and ``R-symmetry'' in the
physical theory.

\item We have referred to the pair $(\CX,\CY)$ as ``defining a vacuum.''  The precise
translation to the physical Coulomb vacuum is described in full detail in \cite{MRV3}.
In the framed case there are physical parameters $(u,\zeta)$, where $u$ is a point on the Coulomb
branch and $\zeta$ is the phase mentioned above, and the mapping to $(\CX,\CY)$ is defined by
\be\label{eq:13}
\begin{split}
\CX & = \Im(\zeta^{-1} a(u)) \\
\CY & = \Im(\zeta^{-1} a_{D}(u)) \\
\end{split}
\ee
where on the RHS we have used standard physical notations.
The fiber of the mapping does not actually determine a unique vacuum: One can rescale $\zeta$ and
$a$ by a common phase.
In the unframed case the same formulae hold with $-\zeta$ replaced by   the phase of the
N=2 central charge $Z_{\gamma}(a)$. The standard collective coordinate supersymmetric
quantum mechanics involves a potential energy term $\sim \parallel \rmG(\CY) \parallel^2$.
Having a ``moduli space with potential'' is in fact a contradiction in terms; one is usually
confined to working in a small potential energy approximation. In \cite{MRV3} it is suggested
that \eqref{eq:13} takes into account all the quantum corrections to the small potential
energy approximation.

\item When hypermultiplet matter in a quaternionic representation $R$ of the gauge
group is included the above definitions are extended
by modifying the Dirac operator so that it also couples to the bundle associated
to the universal bundle by the representation $R$. This bundle is
endowed with a hyperholomorphic connection  \cite{Manton:1993aa,Gauntlett:1995fu,Brennan}.

\end{enumerate}

\subsection{Building BPS States From $\CK_0$}

A natural question to ask, and one which is important to our
discussion of Fredholm conditions below, is whether spinors in the space $\CK_0$
defined in \eqref{eq:CK0} can be used to build BPS states
in \eqref{eq:BPSdef}.
In this section we describe those subspaces of $\CK_0$.
Accordingly, let  us begin with the Dirac operator $\Dsl_0^{\rm Y}$ on $\CM_0(\gamma_{\mathrm{m}}, \CX)$
for an element ${\rm Y} \in \ft_0$.
Only the subspace $\ft_0\subset \ft$ orthogonal to $\gamma_{\mathrm{m}}$ in the Killing metric
will act effectively on $\CM_0(\gamma_{\mathrm{m}}, \CX)$ so our question becomes:
When are spinors in the $\ft_0$-isotypical components of $\CK_0$
factors of BPS spinors $\Psi \in \CH(\gamma; \CX, \CY)$ for some $\gamma^{\mathrm{e}}\in \Lambda_{\mathrm{rt}}$ and $\CY\in \ft$?

In order to answer this question introduce the
rank one projection operator $\textbf{Q}: \ft^\vee \to \ft^\vee$ dual to $\textbf{P}$,
i.e.   $\langle \textbf{Q}(\gamma^{\mathrm{e}}), h \rangle :=\langle \gamma^{\mathrm{e}}, \textbf{P}(h) \rangle $
for all $h \in \ft$. Then one computes $\textbf{Q}(\gamma^{\mathrm{e}}) = \frac{\langle \gamma^{\mathrm{e}}, \CX \rangle}{(\gamma_{\mathrm{m}}, \CX)} \gamma_{\mathrm{m}}^\vee$.
Since $\textbf{Q}$ is  a projection operator  we can use it to give a decomposition
\be
\ft^\vee \cong  {\rm Im}\textbf{Q}   \oplus {\rm Im}(1-\textbf{Q})
:= \ft_{\mathrm{com}}^\vee \oplus \ft_0^\vee
\ee
The isotypical decomposition of the $\ft_0$ action on $\CK_0$ is a sum over
characters $\gamma^{\mathrm{e}}_0\in \ft_0^\vee$.  However, we can only complete a spinor $\Psi_0\in \CK_0$
with a spinor $\Psi_{\mathrm{com}} \in \CK_{\mathrm{com}}$ if the electric charge is properly quantized,
and the condition for this is
\be\label{eq:ge0-quant}
\gamma^{\mathrm{e}}_0 \in (1-\textbf{Q})(\Lambda_{\mathrm{rt}})  ~ .
\ee
Let $\CL(\gamma_0^{\mathrm{e}}) := \Lambda_{\mathrm{rt}} \cap (1 - \textbf{Q})^{-1}(\gamma_0^{\mathrm{e}}) $.
For each choice of $\gamma^{\mathrm{e}} \in  \CL(\gamma_0^{\mathrm{e}})$    compute  $q$ via
\eqref{eq:Defq}, and form $\Psi = \Psi_{\mathrm{com}}\otimes \Psi_0$.
Moreover, we define ${\rm Y}' \in \ft_{\mathrm{com}}$ to be
the solution to   \eqref{eq:Compat} and
set $\CY := {\rm Y}' + {\rm Y}$. We now have the quartet
$(\CX,\CY,\gamma_{\mathrm{m}}, \gamma^{\mathrm{e}})$ and $\Psi\in \CK^{\gamma^{\mathrm{e}}}$ potentially
describes a BPS state. However, we still must require that  $\Psi$ descend
from $\widetilde{\CM}$ to $\CM$. Recall that $\phi^*(x^4) = x^4 + 2\pi$
and therefore we must require that
\be\label{eq:Psi0-equiv}
\phi^*(\Psi_0) = e^{ 2\pi \I q} \Psi_0.
\ee

Now that we know when we can build BPS states from elements of $\CK_0$ we should
ask about uniqueness.
 Suppose that we have found an element of  $ \CL(\gamma_0^{\mathrm{e}})$
so that \eqref{eq:Psi0-equiv} holds. What other charges in  $\CL(\gamma_0^{\mathrm{e}})$
will also support BPS states? Note that $\CL(\gamma_0^{\mathrm{e}}) $
is a torsor for
$\Lambda_{\mathrm{rt}} \cap \ker (1- \textbf{Q})= \Lambda_{\mathrm{rt}} \cap \Im\textbf{Q}$
and since $\Im\textbf{Q}$ is just the real line through $\gamma_{\mathrm{m}}^\vee$ this
is simply the set of multiples (necessarily rational) of $\gamma_{\mathrm{m}}^\vee$ that are
in the root lattice. Therefore
  $\CL(\gamma_0^{\mathrm{e}})\subset \Lambda_{\mathrm{rt}}$ is a $\IZ$-torsor where the $\IZ$-action
is generated by  $\gamma^{\mathrm{e}} \to \gamma^{\mathrm{e}} + \frac{1}{\ell}\gamma_{\mathrm{m}}^\vee$, that is,
any two elements of the torsor are related by shifting
\be\label{eq:k-shift}
 \gamma^{\mathrm{e}} \to \gamma^{\mathrm{e}} + \frac{k}{\ell} \gamma_{\mathrm{m}}^\vee
\ee
where $k$ is an integer. By \eqref{eq:Defq} this  has the effect of shifting:
$q \to q + \frac{k}{\ell}$.
Therefore, if \eqref{eq:Psi0-equiv} has a solution for some $\gamma^{\mathrm{e}}$
and we shift by integral multiples of $\gamma_{\mathrm{m}}^\vee$, that is, if $k$ is
a multiple of $\ell$,
then the same spinor $\Psi_0\in \CK_0^{\gamma^{\mathrm{e}}_0}$ will satisfy \eqref{eq:Psi0-equiv} and hence we
can use the same $\Psi_0$, but a different $\Psi_{\mathrm{com}}$, to construct a new BPS state.
We call the collection of BPS states thus obtained a \emph{Julia-Zee tower}.
On the other hand, if we consider a shift \eqref{eq:k-shift} where $k$ is not a multiple
of $\ell$ then the equivariance condition \eqref{eq:Psi0-equiv} changes. There might, or might not,
be a different spinor in $\CK_0^{\gamma^{\mathrm{e}}_0}$ transforming according to this new equivariance
condition.

\section{Mathematical Predictions}\label{sec:Predictions}

In this section we state three mathematical conjectures
following from recent results in the physics literature.

\subsection{Predictions From The No-Exotics Theorem}

As we have remarked, the semiclassical spaces of
framed and unframed BPS states:
$$\overline{\underline{\CH}}([P] ; \gamma;\CX,\CY),$$
and
$\CH_0(\gamma;\CX,\CY)$, respectively, are
representations of $\fs\fo(3)_{\rm rot} \oplus \fs\fu(2)_{R}$.
The ``no exotics conjecture'' of \cite{Gaiotto:2010be} states that
 $\fs\fu(2)_R$ should act trivially on these spaces:
They are direct sums of $\mathrm{SU}(2)_R$ singlets. This has been proven  for gauge groups SU$(K)$
(using
the relation of BPS states to the cohomology of spaces of
representations of quivers and the Kontsevich-Soibelman wall-crossing
formula)
in
\cite{Chuang:2013wt} and for all simply laced gauge groups in
\cite{DelZotto:2014bga}. Recently a very beautiful general argument
based on the structure of supersymmetric stress-tensor multiplets
has been given \cite{CordovaDumitrescu}.   To see one simple implication for the
Dirac operator we note that under the homomorphism
\be\label{eq:15}
{\rm SU}(2)_R \times {\rm Sp}(N)_{\rm hol}  \rightarrow {\rm Spin}(4N)
\ee
the element $(-1,1)$ maps to the Clifford volume form with
the natural orientation provided by any of the complex structures on
the moduli space. It therefore follows that all the spinors in the
kernel of the Dirac operator have positive chirality.

Moreover, choosing any complex structure on the moduli space
we can express the no-exotics condition in terms of the Dolbeault
operator acting on sections of $\Lambda^{0,*}$:
\be\label{eq:16}
\bar\p^{\CY}:= \bar \p + \rmG(\CY)^{0,1}\wedge
\ee
using the isomorphism of the spinor bundle with $\Lambda^{0,*}$
\cite{HitchinHarmonicSpinors}. In this formulation a set of
generators of $\fs\fu(2)_{R}$ can be chosen so that
\be
\begin{split}
I^3\vert_{\Lambda^{0,q} } & = \half (q-N) \textbf{1} \\
I^+ = \upomega^{0,2} \wedge \qquad & \qquad I^- = \iota(\upomega^{2,0)}) \\
\end{split}
\ee
By the no-exotics theorem these operators must act trivially on
the cohomology and hence the $L^2$-cohomology of
$H^{0,q}$ with respect to $\bar\p^{\CY}$ (or at least that part
satisfying the equivariance conditions with respect to the
Deck transformation, as discussed above)  is concentrated in
the middle degree, $q=N$, and is primitive in the sense that
it is annihilated by product with the  $(0,2)$ combination of K\"ahler forms
or by contraction with the   $(2,0)$ combination of K\"ahler forms.

In ongoing work \cite{Brennan} this discussion is generalized to
include gauge theories with hypermultiplet matter.
The $L^2$-cohomology of the corresponding operator $\bar\p^{\CY}$
is still primitive and concentrated in the middle degree.
In particular, specializing to  $\fg = \fs\fu(2)$  with a hypermultiplet
in the adjoint   we obtain precisely the renowned prediction
of Ashoke Sen \cite{Sen:1994yi} based on $S$-duality.

\subsection{Predictions From Wall-Crossing Formulae}

We now consider the families of Dirac operators as we vary the
Higgs fields. For simplicity our discussion will focus on the family parametrized by $\CY \in \ft$
in the framed case, and by ${\rm Y} \in \ft_0$ in the unframed case.
\footnote{One can  also let $\CX$ vary in the fundamential Weyl chamber.
Strictly speaking, in order to vary $\CX$ we should define a
connection on the bundle of hyperk\"ahler manifolds over the chamber.
This is not worked out in \cite{MRV3}, but we trust it is easily taken care of.}
Generically, $\Dsl^{\CY}$ acting on $L^2$ spinors
over $\froM(P,\gamma_{\mathrm{m}}, \CX)$ (and $\Dsl_0^{\rm Y}$ acting on $L^2$
spinors over $\CM_0(\gamma_{\mathrm{m}};\CX)$) is  Fredholm,
and by the no-exotics property, the character of $\ker \Dsl_+^{\CY}$ is the same as the
character-valued index. Therefore, we expect the $\fs\fo(3)$ characters of
$\overline{\underline{\CH}}([P]; \gamma;\CX,\CY)$
and $\CH_0(\gamma;\CX,\CY)$ to be piecewise constant
as functions of $\CY\in \ft$. However, on real codimension one walls, 
$\Dsl^{\CY}$, respectively  $\Dsl_0^{\rm Y}$,  will fail to be Fredholm.
\footnote{Our statements generalize the discussion in \cite{Stern:2000ie}.}
Physically, these walls
correspond to walls of marginal stability, and using standard formulae for
the $N=2$ central charges we are then led to the following predictions:

First, a translation of the standard criterion for crossing
 a wall of marginal stability, taking into account simplifications
 from working in the semiclassical limit, leads to the following
 criterion:   $\Dsl_0^{\rm Y}$ is not Fredholm only if there are charges $\gamma_1, \gamma_2\in \Gamma$,
 with $\gamma_{1,\mathrm{m}} + \gamma_{2,\mathrm{m}}= \gamma_{\mathrm{m}}$,
``nonlocal'' in the sense that   $\llangle \gamma_1, \gamma_2 \rrangle \not=0$, and ``occupied'' in
the sense that  $\CH(\gamma_i; \CX,\CY) \not= 0 $, for both $i=1,2$, where $\CY$ is constructed
from  ${\rm Y}$ and a choice of $\gamma^{\mathrm{e}} = \gamma^{\mathrm{e}}_1 + \gamma^{\mathrm{e}}_2$ as
described above. Note that \eqref{eq:Compat} implies that
\be
(\gamma_{i,\mathrm{m}}, \CY) + \langle \gamma_{i}^{\mathrm{e}}, \CX \rangle =0, \qquad \qquad i=1,2,
\ee
but these two equations are not independent of each other because their
sum gives the identity \eqref{eq:Compat} for the total charge $\gamma=\gamma_1 + \gamma_2$.
Indeed the  the codimension wall on which $\Dsl_0^{\rm Y_0}$ will fail to be
Fredholm (due to this pair of charges $\gamma_1, \gamma_2$ ) is defined by 
(either) one of the above conditions, which can be written in the form: 
\be\label{eq:17}
({\rm Y}, \gamma_{\mathrm{m},1}) + \langle \gamma^{\mathrm{e}}_1 - \gamma_{\mathrm{m},1} \frac{ \langle \gamma^{\mathrm{e}}, \CX \rangle}{(\gamma_{\mathrm{m}}, \CX)}, \CX \rangle =0.
\ee
Equation
\eqref{eq:17} defines a real codimension one wall in $\ft_0$.
If ${\rm Y} $ crosses this wall the   space $\CH(\gamma;\CX,\CY)$ will jump
in a way dictated by the   Kontsevich-Soibelman wall-crossing formula \cite{Kontsevich:2008fj}.
We expect that there is a relatively simple physical
argument that shows that the continuous spectrum of $\Dsl_0^{\rm Y}$ extends down to
zero at these real codimension one walls. It entails an analysis of the
square of the operator $\Dsl_0^{\rm Y}$ using
 the asymptotic form of moduli space metrics discussed in
\cite{Gibbons:1995yw,Weinberg:2006rq}. We hope to address this elsewhere.

Second, for singular monopoles,  $\Dsl^{\CY}$ is   Fredholm
except on real codimension one walls defined by
\be\label{eq:18}
W(\gamma_{\mathrm{h}}) := \{ \CY:  (\gamma_{\mathrm{h},\mathrm{m}}, \CY) + \langle \gamma_{\mathrm{h},\mathrm{e}} , \CX \rangle =0 \quad \& \quad
\CH(\gamma_{\mathrm{h}}; \CX,\CY) \not=0 \quad \& \quad Z_{\gamma_{\mathrm{h}}}(\zeta^{-1}a) < 0 \}
\ee
(The last condition $Z_{\gamma_{\mathrm{h}}}(\zeta^{-1}a) < 0$ is satisfied automatically in the semiclassical region.)
In this case a very explicit formula for the jump of $\overline{\underline{\CH}}([P],\zeta; \gamma;\CX,\CY)$
follows from \cite{Gaiotto:2010be}: We introduce the Heisenberg group extension of $\IC[\Gamma]$ defined by
$X_{\gamma_1} X_{\gamma_2} = y^{\llangle \gamma_1,\gamma_2\rrangle} X_{\gamma_1+\gamma_2}$ and form the
generating function
\be\label{eq:19}
F([P];\CX,\CY):= \sum_{\gamma\in \Gamma} \fro([P],\gamma;\CX,\CY) X_\gamma
\ee
where $\fro(P,\gamma;\CX,\CY)$  is the $\fs\fo(3)$ character of $\overline{\underline{\CH}}([P]; \gamma;\CX,\CY)$,
that is, the trace of $y^{2J}$ where $J$ is any generator of $\fs\fo(3)$ normalized to have half-integer eigenvalues.
In a similar way, define a set of integers $a_m(\gamma_{\mathrm{h}})$ by the character of $\CH(\gamma_{\mathrm{h}};\CX,\CY)$,
$\Omega(\gamma_{\mathrm{h}};\CX,\CY) = \sum_m a_m (-y)^m$. Then, crossing a wall $W(\gamma_{\mathrm{h}})$ we have the
transformation law $F \to   S F S^{-1}$,
where $S = \prod_{n>0} S_{n \gamma_{\mathrm{h}}}$,
\be\label{eq:21}
S_{\gamma_{\mathrm{h}}} = \prod_m \Phi((-y)^m X_{\gamma_{\mathrm{h}}})^{a_m(\gamma_{\mathrm{h}})}
\ee
and  $\Phi(X) = \prod_{k=1}^\infty (1+ y^{2k-1} X)^{-1}$ is the quantum dilogarithm.

\section{Two Examples}\label{sec:Examples}

\subsection{Example 1: Wall Crossing For Smooth Monopoles}

\begin{figure}[htp]
\centering
\includegraphics[scale=0.8]{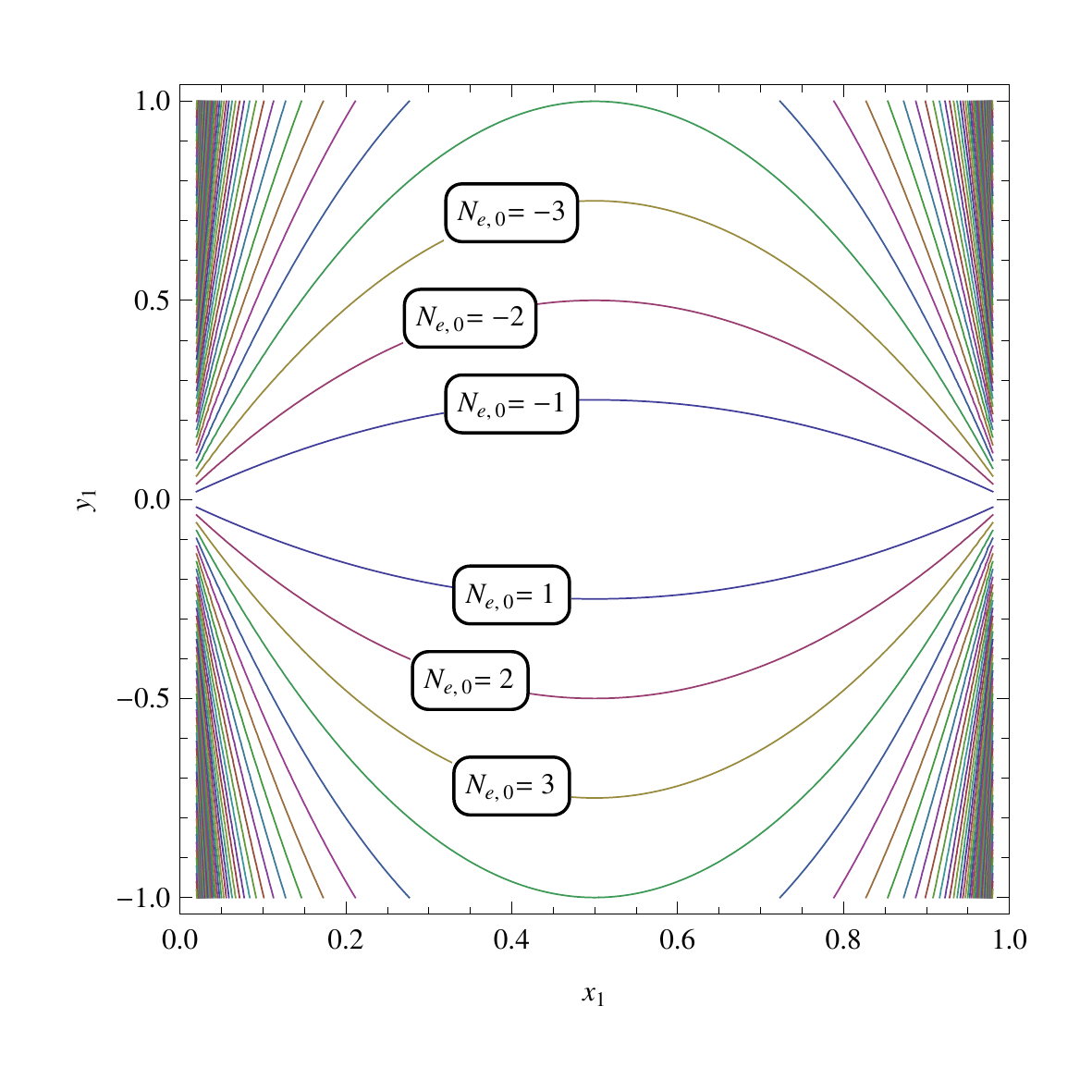}
\caption{Walls of marginal stability for $\fg = \fs\fu(3), \fs\fo(5), \fg_2$ for
$\sp=1,2,3$ respectively. We choose axes to be the scale-invariant quantities
$x_1 := (H_1,\CX)/(\gamma_{\mathrm{m}}, \CX)$ and $y_1 := \sp (H_1, \CY_0)/(\gamma_{\mathrm{m}}, \CX)$.
The chamber containing the line $y_1 = 0$ has no BPS states. Crossing a wall
labeled by $N_{\mathrm{e},0}$ in the direction away from the central chamber adds a tower of BPS states with electric
charge $\gamma^{\mathrm{e}} = n \alpha_1+ (\sp n-N_{\mathrm{e},0})\alpha_2$ with $n\in \IZ$ to the spectrum.
The $\fs\fo(3)_{\rm rot}$ representation
$\CH_0(\gamma;\CX,\CY)$ gains a spin $j$ representation
with $j = \half ( \vert N_{\mathrm{e},0}\vert -1)$. At the walls $x \to 0,1$ the classical 
mass of the monopoles goes to zero and the semiclassical analysis is not necessarily valid.     }
\label{fig:TaubNutWalls}
\end{figure}

We consider the case when $\fg$ has rank $2$, so there are two simple
coroots and $\gamma_{\mathrm{m}} = H_1 + H_2$. What follows is a reinterpretation
of the discussion of \cite{Gauntlett:1999vc}. In this case the
strongly centered moduli space $\CM_0$ is the Taub-NUT metric.
We can represent it as projection $\pi: \CM_0 \to \IR^3$ where the
fiber is a circle, parametrized by $\uppsi \sim \uppsi + 2\pi$, except
for the fiber over the origin, which is just a point. Splitting
$\rmG(\CY)$ to extract the component along $\CM_0$ yields
\be\label{eq:22}
\rmG(\CY_0) =   \frac{1}{\sp} (H_1, \CY_0)\frac{\p}{\p \uppsi}
\ee
where $\sp=1,2,3$ for $\fg = \fs\fu(3),\fs\fo(5), \fg_2$, respectively and we take $\alpha_1$
to be the long root. Electric charges of BPS states are of the form
\be\label{eq:23}
\gamma^{\mathrm{e}} = n_{\mathrm{e}}^1 \alpha_1 + n_{\mathrm{e}}^2 \alpha_2
\ee
with $n_{\mathrm{e}}^1, n_{\mathrm{e}}^2 \in \IZ$. Now $\ft_0$ is one-dimensional and the projection in \eqref{eq:ge0-quant} is
\be
\gamma^{\mathrm{e}}_0 = (\sp n_{\mathrm{e}}^1 - n_{\mathrm{e}}^2) \cdot \left(\frac{ m_2 \alpha_1 - m_1 \sp \alpha_2}{\sp(m_1+m_2)} \right)
\ee
where $m_i := (H_i, \CX)$. Define $N_{\mathrm{e},0} := \sp n_{\mathrm{e}}^1 - n_{\mathrm{e}}^2$. Then, for fixed
$\gamma^{\mathrm{e}}_0$ the preimage of compatible electric charges is the $\IZ$ torsor
with fixed value of $N_{\mathrm{e},0}$, that is the set of electric charges
$\gamma^{\mathrm{e}} = n \alpha_1 + (n \sp - N_{\mathrm{e},0})\alpha_2$ with $n\in \IZ$.
Elements of $\CK_0$ that can be used to build states of definite
electric charge
are equivariant in the $\uppsi$ direction: $ L_{\frac{\p}{\p\uppsi}}\Psi_0 = - \I N_{\mathrm{e},0} \Psi_0 $.
The non-Fredholm condition is now based on a pair of electromagnetic charges
 $\gamma_i =  H_i \oplus n_{\mathrm{e}}^i \alpha_i$,  $i=1,2$ with
 $\llangle \gamma_1, \gamma_2 \rrangle = - N_{\mathrm{e},0} \not=0$.
 The spaces $\CH_0(\gamma_i;\CX,\CY)$ correspond to the case $\CM_0={\rm pt}$ and are one-dimensional
 so the non-Fredholm criterion is satisfied. The jump in the $L^2$ kernel can be verified
 by direct computation in this case since the determination of the
kernel of the Dirac-like operator $\Dsl_0^{\rm Y}$ on $\CM_0$ is a computation
going back to \cite{Pope:1978zx}. The result is illustrated in
Figure \ref{fig:TaubNutWalls}  and is simply the result predicted by
the primitive wall-crossing formula of \cite{Denef:2007vg,Diaconescu:2007bf},
which suffices in this case since the $\gamma_i$ are primitive.
Moreover, the boundstate wavefunction, as a function of the (suitably rescaled) radial direction
$r$ in $\IR^3$, is proportional to $r^{(\vert N_{\mathrm{e},0} \vert -1)/2} e^{-\vert 2C + N_{\mathrm{e},0} \vert r} $
where $\rmG(\CY_0) = \frac{2\mu}{\sp^2} C \frac{\p}{\p \uppsi}$, $\mu^{-1} = m_1^{-1} + m_2^{-1}$, and the
constant $C$ is proportional to the coordinate $y_1$ in Figure \ref{fig:TaubNutWalls}.
The wavefunction is peaked  at $r_{\rm ext}= \vert N_{\mathrm{e},0} \vert/|2C + N_{\mathrm{e},0}\vert $. This turns out to be precisely
the bound-state radius computed from explicit BPS field configurations by
Denef \cite{Denef:2002ru}.

\subsection{Example 2: Wall-Crossing For Singular Monopoles}

We now consider $\fg = \fs\fu(2)$, but consider a singular monopole with
$P = \frac{p}{2} H_\alpha$. We can take $p>0$.
Splitting $\gamma= \gamma_{\mathrm{m}} \oplus \gamma^{\mathrm{e}} $ with $\gamma_{\mathrm{m}} = (\tilde n_{\mathrm{m}} - \half p) H_\alpha$,
we see that $\fro(P,\gamma;\CX,\CY)$ is the $\fs\fo(3)$ character of the Dirac operator
on a moduli space of singular monopoles of dimension $4 \tilde n_{\mathrm{m}}$. The spectrum
of the ordinary BPS states is well-known \cite{Seiberg:1994rs}: With our choice $p>0$ the only relevant
spaces are $\CH_0(\gamma_n;\CX,\CY)\cong \IC$ with $\gamma_n = H_\alpha+ n \alpha$. We can thus divide
up the $\CY$-line into chambers separated by walls $W_n= W(\gamma_n)$. In the chamber
between walls $W_n$ and $W_{n+1}$ the generating function \eqref{eq:19} is given by
\be\label{eq:25}
F_n =\left[ X_1^{-1/2}X_2^{-n/2}\left( U_{n}(f_{n})- X_2^{-1/2}U_{n-1}(f_{n})\right)\right]^{p}
\ee
where $X_1 = X_{H_\alpha}$ and $X_2 = X_\alpha$ generate the Heisenberg extension of $\IC[\Gamma]$,
\begin{equation}\label{eq:26}
f_n:=\half\left( X_2^{1/2}+  X_2^{-1/2}\left(1+y^{2n+3 }X_1X_2^{n+1}\right) \right),
\end{equation}
and $U_n(\cos\theta) = \sin((n+1)\theta)/\sin\theta$. The result is proven
in \cite{MRV3} by careful application of the wall crossing formula for $F([P];\CX,\CY)$  and generalizes the result in
\cite{Gaiotto:2010be} at $y=1$. Equation  \eqref{eq:25} makes some
fairly nontrivial predictions for $L^2$-kernels of Dirac operators on singular monopole
moduli spaces, some of which are spelled out in detail in \cite{MRV3}.

\section*{Acknowledgements}

GM thanks N. Hitchin, C. LeBrun, N. Manton and E. Witten for useful remarks, and D. Brennan for collaboration on related matters.
DVdB is partially supported by TUBITAK grant 113F164 and by the Bo\u{g}azi\c{c}i University Research Fund under grant number 13B03SUP7, GM is supported by the U.S. Department of Energy under grant DOE-SC0010008 to Rutgers University, and ABR is supported by the Mitchell Family Foundation.  GM also gratefully acknowledges
the hospitality  the Isaac Newton Institute; ABR thanks the NHETC at Rutgers University.

\end{document}